\documentclass[]{aa}
\usepackage{graphicx}
\usepackage{adjustbox}
\usepackage{txfonts}
\usepackage{url}
\makeatletter
\renewcommand*\aa@pageof{, page \thepage{} of \pageref*{LastPage}}
\makeatother
\usepackage{hyperref}
\hypersetup{
    colorlinks=true,
    linkcolor=blue,
    filecolor=magenta,      
    urlcolor=blue,
    citecolor=blue,
    }
\begin{document} 

\title{The Small Magellanic Cloud through the lens of the {\it James Webb Space Telescope}: binaries and mass function within the galaxy outskirts}
\subtitle{}

\author{M. V. Legnardi\inst{1} \and F. Muratore\inst{1} \and A. P. Milone\inst{1,2} \and G. Cordoni\inst{3} \and T. Ziliotto\inst{1} \and E. Dondoglio\inst{2}  \and A. F. Marino\inst{2} \and A. Mastrobuono-Battisti\inst{1} \and E. Bortolan\inst{1} \and E. P. Lagioia\inst{4} \and M. Tailo\inst{2} 
}

\institute{Dipartimento di Fisica e Astronomia ``Galileo Galilei'', Univ. di Padova, Vicolo dell'Osservatorio 3, Padova, IT-35122 \\ \email{mariavittoria.legnardi@unipd.it}
\and
Istituto Nazionale di Astrofisica - Osservatorio Astronomico di Padova, Vicolo dell’Osservatorio 5, Padova, IT-35122
\and 
Research School of Astronomy and Astrophysics, Australian National University, Canberra, ACT 2611, Australia
\and
South-Western Institute for Astronomy Research, Yunnan University, Kunming, 650500 P. R. China}
\titlerunning{The binary fraction and mass function of the SMC field} 
\authorrunning{Legnardi et al.}

\date{Received 4 July 2025 / Accepted 8 September 2025}

\abstract{The stellar initial mass function (IMF) and the fraction of binary systems are fundamental ingredients that govern the formation and evolution of galaxies. Whether the IMF is universal or varies with environment remains one of the central open questions in astrophysics. Dwarf galaxies such as the Small Magellanic Cloud (SMC), with their low metallicity and diffuse star-forming regions, offer critical laboratories to address this issue. In this work, we exploit ultra-deep photometry from the {\it James Webb Space Telescope} to investigate the stellar populations in the field of the SMC. Using the $m_{\rm F322W2}$ versus\,$m_{\rm F115W}-m_{\rm F322W2}$ color-magnitude diagram (CMD), we derive the luminosity function and measure the fraction of unresolved binary systems. We find a binary fraction of $f_{\rm bin}^{q>0.6}=0.14\pm0.01$, consistent with results from synthetic CMDs incorporating the metallicity distribution of the SMC. Additionally, the measured binary fraction in the SMC field is consistent with those observed in Galactic open clusters and Milky Way field stars of similar ages and masses, suggesting similar binary formation and evolutionary processes across these low-density environments. By combining the luminosity function with the best-fit isochrone, we derive the the mass function (MF) down to $0.22\,M_{\odot}$, the lowest mass limit reached for the SMC to date. The resulting MF follows a power-law with a slope of $\alpha=-1.99\pm0.08$. This value is shallower than the canonical Salpeter slope of $\alpha=-2.35$, providing new evidence for IMF variations in low-metallicity and low-density environments.}

\maketitle

\section{Introduction}
\label{sec:intro}

The Small Magellanic Cloud (SMC) provides a unique and valuable laboratory for studying stellar populations in a low-metallicity, extragalactic environment. Its proximity, low foreground extinction, and relatively low crowding compared to more massive galaxies enable us to resolve individual stars across a broad range of stellar masses, from low-mass main sequence (MS) stars to evolved giants. As a result, the SMC has become a cornerstone for testing stellar evolution models and investigating the processes that shape galaxies beyond the Milky Way \citep[e.g.,][]{dobbie2014a,dobbie2014b,rubele2018,mackey2018,pastorelli2019}. Among the fundamental ingredients that regulate star formation and evolution are the initial mass function (IMF) and the fraction of binaries, both of which influence the dynamical evolution of stellar systems, the rate of supernovae and chemical enrichment, and the production of exotic objects such as blue stragglers, X-ray binaries, and gravitational wave sources.

Despite their importance, both the IMF and the binary fraction remain poorly constrained in external galaxies, particularly in the field populations of low-mass systems like the SMC. While studies of young clusters in the Magellanic Clouds have provided valuable insights into the high-mass IMF and short-period massive binaries \citep[e.g.,][]{dunstall2015,sabbi2016,almeida2017,schneider2018,kalari2018}, much less is known about the binary properties of field stars, especially at intermediate and low masses. In this regime, unresolved binaries can significantly bias luminosity and mass function (MF) estimates by mimicking brighter, higher mass single stars in the color-magnitude diagram (CMD). Properly accounting for binary contamination is therefore crucial for deriving accurate stellar MFs.

The interplay between the IMF and binaries is especially relevant in the context of the ongoing debate about the universality of the IMF, that is, whether its shape remains constant across different environments and over cosmic time. While the classical IMF by \citet{salpeter1995} describes a power-law distribution with slope $\alpha = -2.35$ over $0.4 \lesssim M/M_\odot \lesssim 10$, numerous studies have suggested deviations from this form. Simulations and ALMA observations point to top-heavy IMFs in early stars and dense star-forming regions \citep[e.g.,][]{abel2002, bromm2002, pouteau2022}, whereas integrated-light studies of giant ellipticals suggest bottom-heavy IMFs dominated by low-mass stars \citep{vandokkum2010, conroy2012}. Stellar kinematic analyses have reached similar conclusions \citep{cappellari2012}.

Although integrated-light studies offer IMF constraints in diverse environments, they are inherently model-dependent. Conversely, resolved stellar populations enable more robust IMF measurements, free from many of these uncertainties \citep[e.g.,][]{luhman2009,milone2012b,sollima2017,dondoglio2022,cordoni2023,baumgardt2023,marino2024a}. However, these studies are mostly confined to the Milky Way and its clusters, limiting the parameter space in which IMF variations can be tested.

Dwarf galaxies like the SMC and ultra-faint dwarfs (UFDs) offer ideal laboratories for probing IMF universality across a broader range of conditions. They differ significantly from the Milky Way in terms of morphology, metallicity, star formation history, and stellar density. Additionally, their extremely long two-body relaxation times (e.g., $\sim 2 \times 10^4$ Gyr in the SMC; \citealt{gennaro2018a}) imply that dynamical effects are negligible, preserving the primordial MF. Observations with the {\it Hubble Space Telescope} ({\it HST}) and the {\it James Webb Space Telescope} ({\it JWST}) have allowed direct star counts in these systems, revealing evidence for a shallower MF slope compared to the canonical value. For example, \citet{wyse2002} found $\alpha = -1.8$ in Ursa Minor, while \citet{geha2013} and \citet{gennaro2018a,gennaro2018b} reported similarly shallow slopes in several UFDs.

In this context, \citet{kalirai2013a} used deep {\it HST} observations with the Wide Field Camera of the Advanced Camera for Surveys (WFC/ACS) to investigate the IMF in the outskirts of the SMC. They reported a best-fit IMF slope of $-1.90^{+0.15}_{-0.10}$, which is shallower than the Salpeter's canonical value. Moreover, their analysis demonstrated that the data is well-represented by a single-component IMF. 

WFC/ACS observations allowed \citet{kalirai2013a} to probe stellar masses down to $\sim 0.4\,M_{\odot}$. In this work, we leverage the exceptional capabilities of {\it JWST} to measure the fraction of blue straggler stars (BSSs) and the binary fraction along the MS, and to investigate the stellar mass distribution in the SMC field down to $\sim 0.2\,M_\odot$. This study provides the first estimate of the photometric binary fraction in the SMC and represents the deepest exploration of the IMF in an extragalactic environment to date. By accounting for binary populations in our analysis, we aim to derive an accurate IMF and assess possible environmental deviations from the IMF observed in the Milky Way and nearby dwarf galaxies.

The paper is organized as follows. In Section~\ref{sec:data}, we describe the photometric dataset and the data reduction procedures. Section~\ref{sec:phot} presents the derivation of the binary fraction in the SMC field. In Section~\ref{sec:MF}, we determine the MF and its slope. Finally, Section~\ref{sec:concl} summarizes our key findings and conclusions.  

\begin{figure*}
    \centering
    \includegraphics[width=.95\textwidth,trim={0cm 6cm 0cm 0cm},clip]{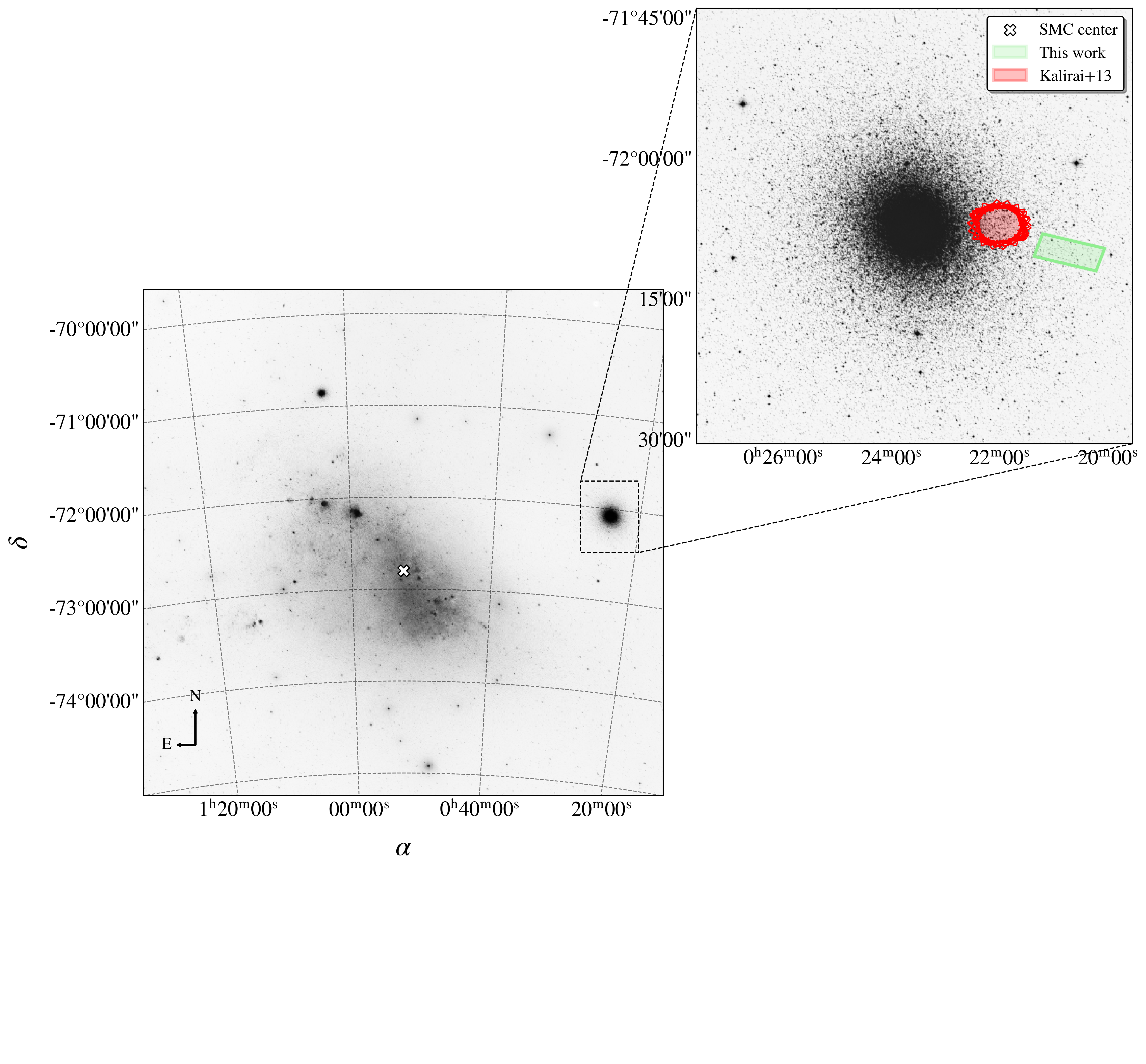}       
    \caption{Image of the SMC obtained from the Digitized Sky Survey 2 (DSS2). The inset provides a close-up view of the region near 47\,Tucanae where the observations for this study were carried out. The NIRCam field of view analyzed in this work is outlined in green, while the footprints of the observations used by \citet{kalirai2013a} are shown in red. The white cross marks the position of the SMC center \citep{piatti2021}. North is up and east is to the left.}
    \label{fig:SMC_data}
\end{figure*}

\section{Observations and data reduction}
\label{sec:data}
To measure the binary fraction and investigate the low-mass end of the stellar MF in the field of the SMC, we used the phtometric catalogs provided by \cite{marino2024a}, which are obtained from images in the F115W and F322W2 filters of the Near-InfraRed Camera (NIRCam) on board the {\it JWST}. These data were collected during the first {\it JWST} observing cycle as part of the GO-2560 program (PI: A. F. Marino). This program is designed to investigate multiple stellar populations among low-mass stars in the globular cluster (GC) NGC\,104 (47\,Tucanae), leveraging the synergy between the Near-InfraRed Spectrograph (NIRSpec) and NIRCam facilities \citep[see][for details on the dataset and on the program]{marino2024a,marino2024b}. Although the primary target of these observations is 47\,Tucanae, the specific field observed, located at $\alpha \sim 00^{\rm h}21^{\rm m}16^{\rm s}$ and $\delta \sim -72^{\rm d}06^{\rm m}16^{\rm s}$, also encompasses a line of sight intersecting the outskirts of the SMC. In particular, the analyzed SMC populations are $\sim 2.5^{\circ}$ (2.7 kpc) west of the galaxy's center. 

The footprints of the data used in this study are shown in green in the inset of Fig.~\ref{fig:SMC_data}. For comparison, we also show in red the footprints of the observations employed by \citet{kalirai2013a} to measure the IMF in the SMC field in the mass range 0.37-0.93\,$M_\odot$. Although their field lies in close proximity to ours, it is centered at $\alpha \sim 00^{\rm h}22^{\rm m}39^{\rm s}$ and $\delta \sim -72^{\rm d}04^{\rm m}04^{\rm s}$ and covers a distinct region. This spatial separation enables us to provide an independent determination of the IMF in the SMC field.   

The methods used to reduce the observations are described in detail in \citet[][see also \citealt{milone2023b}]{marino2024a}. Briefly, the measurement of stellar fluxes and positions involved a two-step process. In the first step, photometry was performed using the program \texttt{img2xym}, originally developed by \citet{anderson2006} to reduce {\it HST} images. This program utilizes a spatially variable point spread function to determine stellar positions and magnitudes from each individual exposure. Magnitudes were standardized to a common photometric zero-point, using the deepest exposure as the reference frame, while geometric distortions were corrected with solutions provided by \citet{milone2023b}.

The second step involved refining photometry and astrometry using the KS2 program, an updated version of \texttt{kitchen\_sync} by \citet{anderson2008}. KS2 incorporates specialized methods tailored to optimize photometry for stars of varying luminosities \citep[see][for details]{sabbi2016,bellini2017,nardiello2018}. It also provides diagnostic parameters to assess photometric and astrometric quality. These parameters were used, following the approach of \citet[][see their Sec.~2.4]{milone2023a}, to select a reliable sample of well-measured stars.

The photometry was calibrated to the Vega system using the procedure described by \citet{milone2023a} and zero-points available on the Space Telescope Science Institute’s webpage\footnote{\url{https://jwst-docs.stsci.edu /jwst-near-infrared-camera/nircam-performance/nircam-absolute-flux-calibration-and-zeropoints}}. Furthermore, reddening variations across the field of view were confirmed to be insignificant \citep{legnardi2023}, so no corrections for differential reddening were necessary.

To assess the uncertainties and the completeness level of our photometry, we performed artificial star (AS) tests following the methodology outlined by \citet{anderson2008}. We generated a catalog of 10$^{5}$ ASs with fixed positions and fluxes, distributing them across the field of view to match the spatial distribution of the real stars. Their magnitudes were assigned based on the fiducial line of the SMC MS in the $m_{\rm F322W2}$ versus\,$m_{\rm F115W}-m_{\rm F322W2}$ CMD. The ASs were reduced using the KS2 program, following the same procedures applied to real stars. We used diagnostic parameters provided by KS2 to assess the quality of the recovered sources.

To quantify the completeness, we binned the ASs into magnitude intervals of 0.3 mag. In each bin, the completeness fraction was calculated as the ratio of recovered to injected stars. We determined that the 50$\%$ completeness limit is reached at $m_{\rm F322W2}=26.22$ mag.

\begin{figure*}
    \centering
    \includegraphics[width=.9\textwidth]{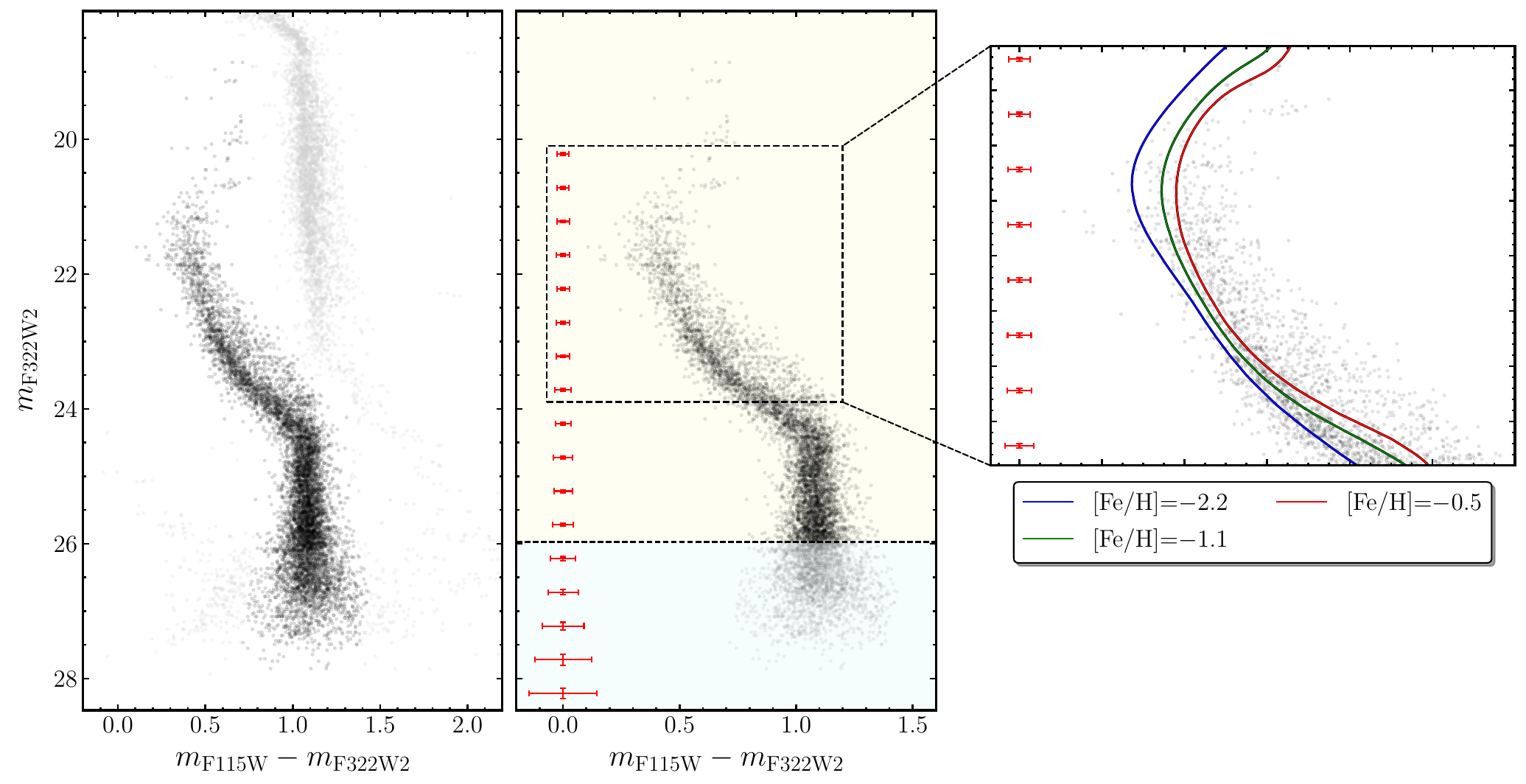}       
    \caption{Illustration of the photometric data used in this work. {\it Left panel.} The $m_{\rm F322W2}$ vs.\,$m_{\rm F115W}-m_{\rm F322W2}$ CMD, showing all detected stars. Likely SMC members and 47\,Tucanae stars are marked with black and gray points, respectively. {\it Right panel.} The same CMD, restricted to stars in the field of the SMC. The black dashed line corresponds to the 50 $\%$ completeness limit for SMC stars in this dataset. The inset zooms in on the upper MS ($20.4 < m_{\rm F322W2} < 23.8$), where three isochrones with an age of 7 Gyr, distance modulus $(m - M)_0 = 18.9$, and reddening $E(B-V) = 0.05$ are overplotted. The blue and red lines correspond to metallicities of [Fe/H] = $-2.2$ and $-0.5$, while the green line indicates the best-fitting isochrone with [Fe/H] = $-1.1$.}
    \label{fig:cmds}
\end{figure*}

\section{The color-magnitude diagram of the Small Magellanic Cloud}
\label{sec:phot}
A visual inspection of the $m_{\rm F322W2}$ versus\,$m_{\rm F115W}-m_{\rm F322W2}$ CMD, shown in the left panel of Fig.~\ref{fig:cmds}, reveals the presence of two distinct stellar populations. The redder sequence (gray points) corresponds to the faint end of the 47\,Tucanae MS, while the bluer sequence (black points) traces field stars belonging to the SMC. In addition, the gray points located in the bottom-left corner of the CMD likely represent the white dwarf cooling sequence of 47\,Tucanae.

The right panel of Fig.~\ref{fig:cmds} isolates the SMC field population, highlighting a significant broadening of the MS in the $m_{\rm F115W}-m_{\rm F322W2}$ color that exceeds expectations from photometric uncertainties alone, as indicated by the error bars on the left corner. This spread is primarily driven by variations in stellar metallicity. Numerous photometric and spectroscopic studies have confirmed the presence of a radial metallicity gradient within the SMC \citep[e.g.,][]{carrera2008,sabbi2009,dobbie2014b,choudhury2020}. For example, \citet{carrera2008} reported that more metal-poor stars tend to dominate at larger galactocentric distances. Similarly, \citet{mucciarelli2023} reported [Fe/H] values from $\sim -2.2$ to $\sim -0.5$ across three fields at increasing distances from the galaxy center\footnote{\citet{mucciarelli2023} analyzed spectra of over 300 stars in three SMC fields centered on the clusters NGC\,121 (FLD-121), NGC\,339 (FLD-339), and NGC\,419 (FLD-419). FLD-121 ($\alpha \sim 00^{\rm h}26^{\rm m}49.0^{\rm s}$, $\delta \sim -71^{\rm d}32^{\rm m}09.9^{\rm s}$), FLD-339 ($\alpha \sim 00^{\rm h}57^{\rm m}48.9^{\rm s}$, $\delta \sim -74^{\rm d}28^{\rm m}00.1^{\rm s}$), and FLD-419 ($\alpha \sim 01^{\rm h}08^{\rm m}17.7^{\rm s}$, $\delta \sim -72^{\rm d}53^{\rm m}02.7^{\rm s}$) are located $\sim 2.5$° north-west, $\sim 1.4$° south-east, and $\sim 1.5$° east of the SMC center, respectively.}, with the fraction of metal-poor stars rising outward.  

In addition to metallicity effects, unresolved binary systems also contribute to the observed broadening of the MS. This is illustrated in the inset of Fig.~\ref{fig:cmds}, which zooms in on the upper MS and displays three BaSTI isochrones \citep{pietrinferni2021} computed for an age of 7 Gyr, distance modulus $(m - M)_0 = 18.9$, and reddening $E(B-V) = 0.05$ but spanning the full metallicity range observed in the SMC field. The blue and red curves represent the extremes of the metallicity distribution reported by \citet{mucciarelli2023} ([Fe/H] = $-2.2$ and $-0.5$), while the green curve corresponds to the best-fit metallicity of [Fe/H] = $-1.1$ \citep{kalirai2013a}. While these metallicity variations introduce some MS broadening, they fall short of reproducing the full extent of the observed spread. In particular, the population of stars lying redward of the fiducial MS is too red to be explained by metallicity alone. This discrepancy points strongly to a significant population of unresolved binaries, whose composite colors shift their positions in the CMD toward redder values, thereby accounting for the excess broadening on the red side of the MS.

In the following, we determine for the first time the binary fraction of the SMC field population. In Sec.~\ref{subsec:bin}, we derive the binary fraction under the assumption of a simple stellar population, while in Sec.~\ref{subsec:binfe}, we account for the effects of metallicity variations on the binary fraction measurement. Furthermore, in Sec.~\ref{subsec:BSSs}, we explore the connection between the identified BSSs and the binary star population.

\begin{figure*}
    \centering
    \includegraphics[width=.9\textwidth]{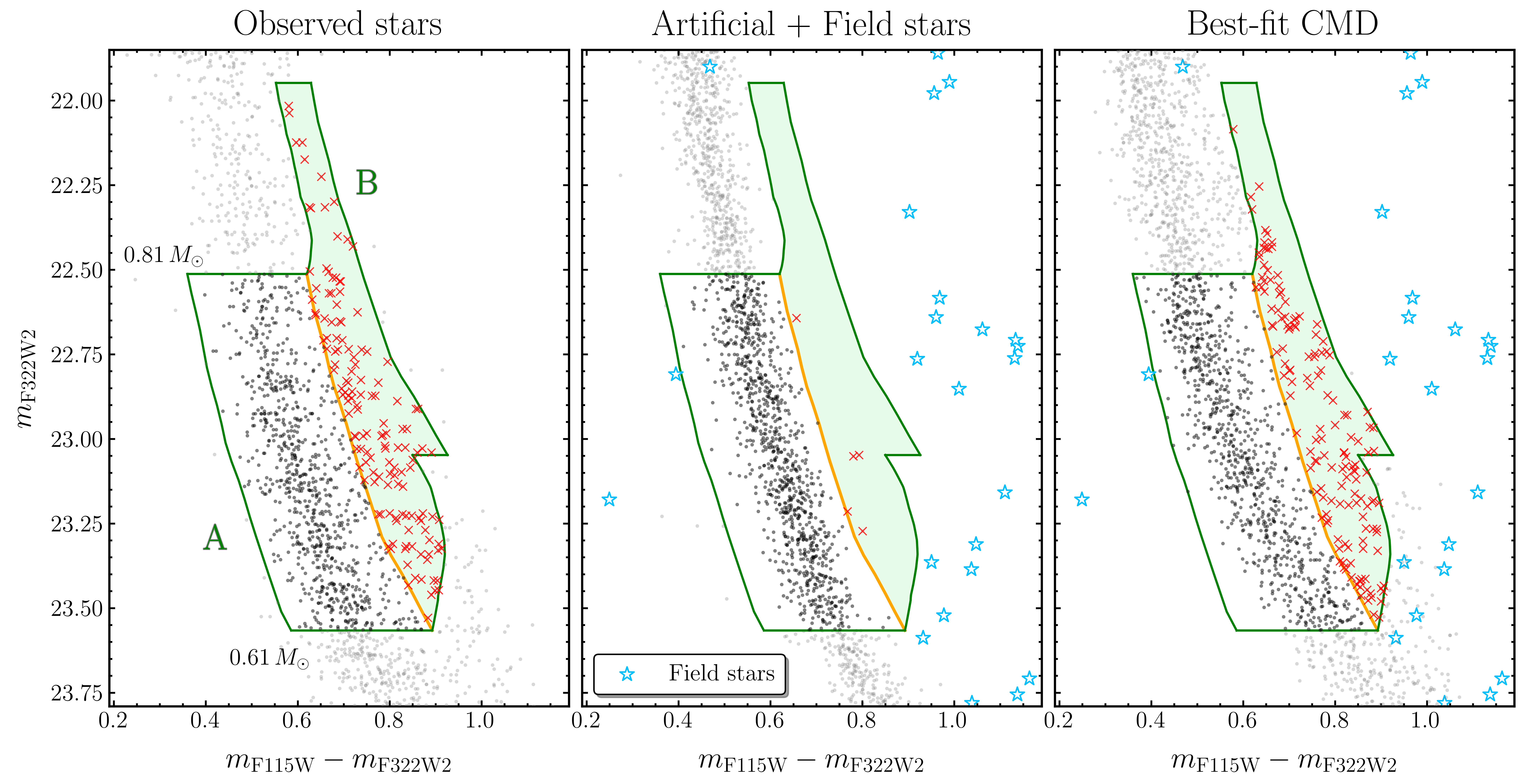}       
    \caption{Binary fraction estimation in the SMC field. {\it Left panel.} $m_{\rm F322W2}$ vs.\,$m_{\rm F115W}-m_{\rm F322W2}$ CMD of SMC stars. Region A (green solid line) includes MS stars and binaries with primary masses between 0.61 and 0.81 $M_{\odot}$. Region B (green-shaded area) contains binaries with $q>0.6$. MS and binary stars are indicated with black points and red crosses, respectively, while the remaining stars are colored in gray. {\it Central panel.} Same as left panel but for artificial and field stars. Azure starred symbols represent stars simulated with the TRILEGAL code \citep{girardi2005} within a Galactic field with the same area and coordinates as the one investigated in this work. {\it Right panel.} The best-fit simulated CMD. Colors and symbols follow the same scheme as in the left and central panels for comparison.}
    \label{fig:bin}
\end{figure*}

\begin{figure}
    \centering
    \includegraphics[width=.45\textwidth]{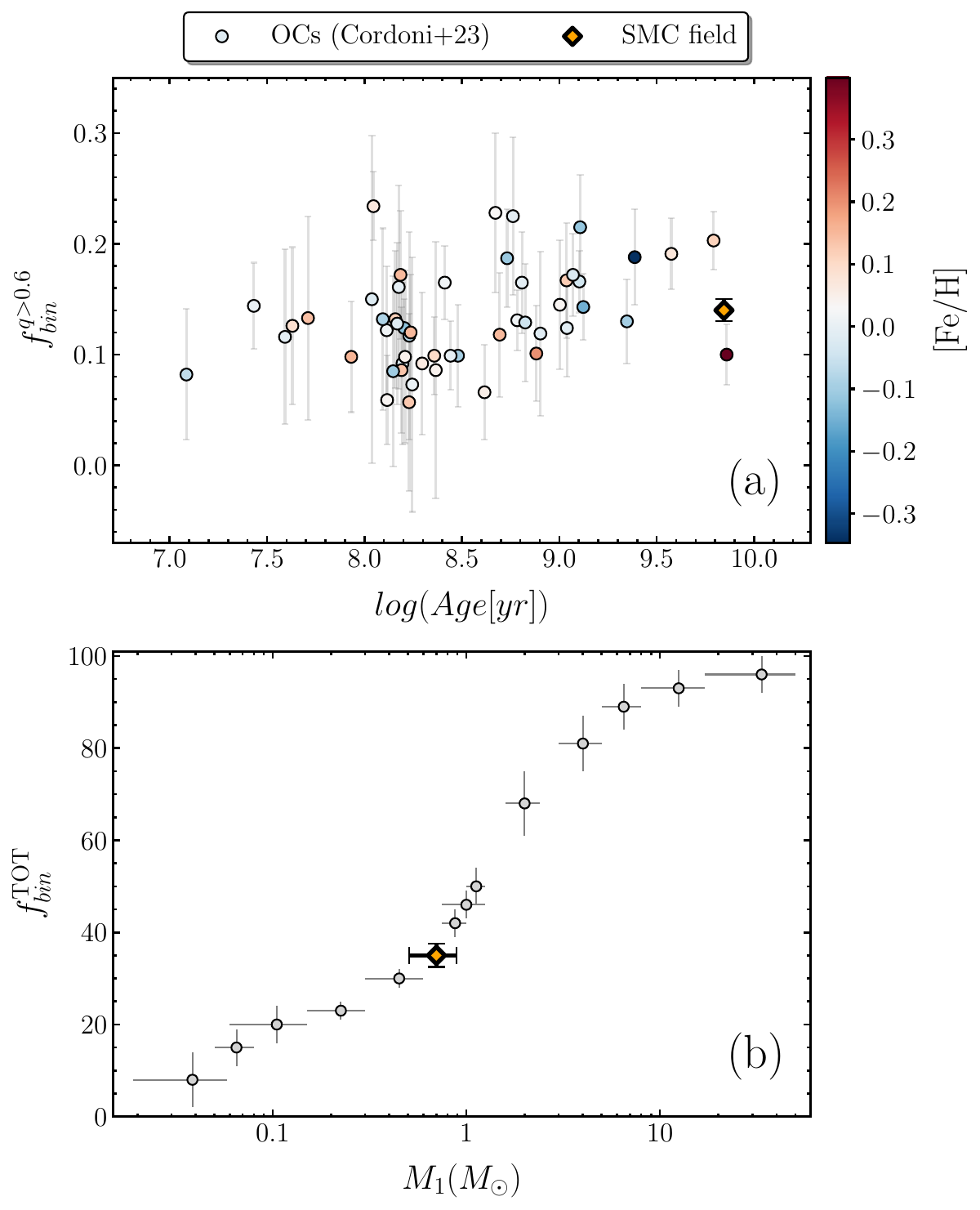}       
    \caption{Comparison of binary fractions in the SMC field and Galactic populations. {\it Panel a.} Binary fraction with $q>0.6$ as a function of the logarithm of cluster age. Galactic OCs are color-coded by their metallicity, as indicated by the color bar on the right, while the orange diamond marks the binary fraction in the field of the SMC. The binary fractions and ages for OCs are from \citet{cordoni2023} and \citet{dias2021}, respectively. {\it Panel b.} Total binary fraction as a function of the mass of the primary star. Gray dots are taken from the review by \citet{offner2023}, while the orange diamond indicates the binary fraction measured for the SMC field in this work.}  
    \label{fig:binplot}
\end{figure}

\subsection{Binaries in the field of the Small Magellanic Cloud}
\label{subsec:bin}
The distance of the SMC, namely $\sim 60.3$ kpc \citep[see also][]{graczyk2020}, prevents the direct resolution of individual components in binary systems, even with space-based telescopes such as the {\it HST} or the {\it JWST}. As a result, unresolved binaries appear as single sources in photometric observations. The total magnitude of an unresolved binary system is given by:
\begin{equation}
    m_{bin} = m_1-2.5\log \left( 1+\frac{F_2}{F_1}\right) 
\end{equation}
where $m_1$ is the magnitude of the primary star, and $F_1$ and $F_2$ are the fluxes of the primary and secondary components, respectively.

In CMDs, unresolved binaries composed of two MS stars appear as point sources that are systematically redder and brighter than single MS stars. Their precise location in the CMD of a simple stellar population depends on the mass of the primary star and the mass ratio of the system, defined as $q = M_2/M_1$. Systems with $q\approx1$, where both stars have nearly equal masses, follow a sequence parallel to the MS but $\sim 0.75$ magnitudes brighter. Conversely, binaries with low mass ratios ($q\sim0$) are nearly indistinguishable from single stars, as their combined flux remains close to the MS fiducial line. These effects are particularly relevant in the $m_{\rm F322W2}$ versus\,$m_{\rm F115W}-m_{\rm F322W2}$ CMD used in this study, where binaries predominantly populate the red and bright side of the MS fiducial sequence.

Over the past few decades, extensive investigations of binary systems in numerous Galactic and extragalactic star clusters have been carried out \citep[e.g.,][]{bolte1992,cool2002,sollima2007,sollima2010,cordoni2023,mohandasan2024,muratore2024,bortolan2025}. In the following, we estimate for the first time the fraction of photometric binary systems in the field of the SMC. To achieve this, we first approximated the SMC CMD as that of a simple stellar population and adapted the methodology introduced by \cite{milone2012b} to our dataset. Our analysis focuses on binaries with mass ratios $q>0.6$, as systems with lower mass ratios are located too close to the MS fiducial line to be reliably distinguished.

As illustrated in Fig.~\ref{fig:bin}, we first identified two distinct regions in the $m_{\rm F322W2}$ versus\,$m_{\rm F115W}-m_{\rm F322W2}$ CMD. Region A (enclosed by the green solid line) includes both single MS stars with $22.5\le m_{\rm F322W2}\le 23.6$ and binary systems where the primary star falls within the same magnitude range. The boundaries of region A are defined as follows: (i) the left boundary is obtained by shifting the MS fiducial line leftward by the average color error; (ii) the right boundary is defined by shifting the equal-mass binary fiducial sequence rightward by twice the color uncertainty; (iii) the upper and lower boundaries correspond to binary systems where the primary star has magnitudes $m_{\rm F322W2}=22.5$ and 23.6, respectively, covering the full range of possible mass ratios from $q=0$ to $q=1$. Region B is a subset of region A that is primarily populated by binaries with $q>0.6$. Its left boundary is marked by the fiducial sequence of binary systems with $q=0.6$ (orange continuous line), which serves as the threshold for our binary selection criteria.

The observed $m_{\rm F322W2}$ versus\,$m_{\rm F115W}-m_{\rm F322W2}$ CMD is presented in the left panel of Fig.~\ref{fig:bin}. Candidate binary systems, identified as sources within region B, are marked with red crosses, while stars in region A are indicated with black points. The fraction of binaries with $q>0.6$ is computed as: 
\begin{equation}\label{eq_bin}
    f_{\rm bin}^{q>0.6}=\frac{N^{B}_{\rm obs}-N^{B}_{\rm field}}{N^{A}_{\rm obs}-N^{A}_{\rm field}}-\frac{N^{B}_{\rm sim}}{N^{A}_{\rm sim}}
\end{equation}
where $N^{A}_{\rm obs}$ and $N^{B}_{\rm obs}$ represent the number of observed stars, corrected for completeness, in region A and B, respectively. $N^{A}_{\rm sim}$ and $N^{B}_{\rm sim}$ refer to the corresponding star counts in the AS CMD, shown in the central panel of Fig.\ref{fig:bin}, while $N^{A}_{\rm field}$ and $N^{B}_{\rm field}$ indicate the number of field stars within the selected regions. To evaluate the contribution of field stars, we compared the observed $m_{\rm F322W2}$ versus\,$m_{\rm F115W}-m_{\rm F322W2}$ CMD with a simulated CMD generated using the TRILEGAL code \citep{girardi2005}. The simulation was performed for a Galactic field with the same area and Galactic coordinates as those of the field analyzed in this work. Simulated field stars are shown with azure starred symbols in the central panel of Fig.~\ref{fig:bin}. The comparison indicates that Milky Way field star contamination of the SMC CMD sequence is negligible. 

The resulting binary fraction in the field of the SMC is $f_{\rm bin}^{q>0.6}=0.14\pm0.01$. The uncertainties in each term of Eq.~\ref{eq_bin} are determined using Poisson statistics, while the total uncertainty in the binary fraction is derived through standard error propagation.

\subsection{Binaries and metallicity distribution}
\label{subsec:binfe}
To assess whether the presence of stellar populations with distinct metallicities affects the measurement of the binary fraction in the SMC field, we compared the observed binary fraction from Sec.~\ref{subsec:bin} with the corresponding fraction obtained from simulated CMDs that account for the metallicity distribution of SMC stars. The synthetic CMDs were generated through an iterative procedure consisting of the following steps.

First, we created a population of single stars by drawing [Fe/H] values from the SMC metallicity distribution derived by \citet{mucciarelli2023}, accounting also for the corresponding [$\alpha$/Fe] relation reported by the same authors. Each star was then assigned a position in the CMD by linearly interpolating isochrones from the BaSTI stellar evolution models \citep{pietrinferni2021} across a grid of metallicities and [$\alpha$/Fe] abundances. Next, we simulated binary systems following the method introduced by \citet{milone2020a}, which has been extensively applied in previous studies \citep[e.g.,][]{muratore2024, bortolan2025, milone2025}. In short, we generated binaries assuming a flat mass-ratio distribution and selected only those falling within the green-shaded region defined for observed stars. We then constructed a grid of simulated CMDs with binary fractions, $f^{q>0.6}_{\rm bin}$, ranging from 0.00 to 1.00 in steps of 0.01, with the corresponding fraction of single stars set to $1-f^{q>0.6}_{\rm bin}$.

The binary fraction of each simulated CMD was measured using the same procedure described in Sec.~\ref{subsec:bin} for real stars. We then compared the simulated values to observations via a $\chi^{2}$ minimization approach. In particular, $\chi^{2}$ has been defined as:
\begin{equation}
    \chi^2=\sum\frac{(n_{obs}-n_{sim})^2}{n_{sim}}
\end{equation}
\noindent where $n_{obs}$ and $n_{sim}$ are the numbers of observed and simulated stars, respectively, that fall within region A and B. 

The $m_{\rm F322W2}$ versus\,$m_{\rm F115W}-m_{\rm F322W2}$ CMD that provides the best match with the observations is illustrated in the right panel of Fig.~\ref{fig:bin}. As in the other panels of Fig.~\ref{fig:bin}, red crosses indicate candidate binaries located in region B (green-shaded area), while black dots represent stars within region A, enclosed by the green solid line.

The best-fitting model, corresponding to the minimum $\chi^{2}$ value, yields a binary fraction of $f^{q>0.6}_{\rm bin}=0.14\pm0.03$. To estimate the associated uncertainty, we generated 1000 synthetic CMDs assuming the best-fit binary fraction. For each simulation, we measured $f^{q>0.6}_{\rm bin}$ using the same approach as for real stars. The final uncertainty was computed as the root mean scatter of the 1000 individual values of $f^{q>0.6}_{\rm bin}$. 

To assess the impact of the adopted metallicity distribution \citep{mucciarelli2023}, we repeated the entire procedure using a simplified metallicity distribution, where [Fe/H] values were uniformly drawn between $-1.4$ and $-1.0$ dex, as in \citet{kalirai2013a}. Synthetic CMDs generated with this alternative distribution yielded a binary fraction of $f_{\rm bin}^{q>0.6} = 0.15 \pm 0.04$, which is statistically consistent with the one derived using the empirical metallicity distribution. This indicates that the assumed metallicity distribution has a negligible effect on the derived binary fraction.

To compare our findings on SMC field stars with those for Galactic OCs and Milky Way field population, we show in Fig.~\ref{fig:binplot} the fraction of binaries with $q>0.6$, $f_{bin}^{q>0.6}$, as a function of the logarithm of the cluster age (panel a) and the total binary fraction\footnote{We extrapolated the value of the total binary fraction, $f_{bin}^{\rm TOT}$, from the $f_{bin}^{q>0.6}$ by assuming a flat mass-ratio distribution over the range $0<q<1$.}, $f_{bin}^{\rm TOT}$, as a function of the primary star mass (panel b). The orange diamond in both panels marks our measurements for the SMC field. Galactic OCs, with binary fractions from \citet{cordoni2023} and ages from \citet{dias2021}, are color coded according to their metallicity \citep{cordoni2023}, while gray dots show Galactic field star binary fractions from \citet[][and references therein]{offner2023}. 

The values of $f_{bin}^{q>0.6}$ in Galactic OCs exhibit a mild correlation with the logarithm of the cluster age and no trend with the cluster metallicity, while the total binary fraction increases monotonically with primary mass. We found that the binary fraction in the SMC field is comparable to that observed in most Galactic OCs. Furthermore, our results align with the trend between total binary fraction and primary star mass reported for Galactic field stars suggesting that binary star formation and evolution in the low-metallicity environment of the SMC proceed similarly to those in low-density environments of the Milky Way.

\subsection{Binaries and blue straggler stars}
\label{subsec:BSSs}
\begin{figure}
    \centering
    \includegraphics[width=.45\textwidth]{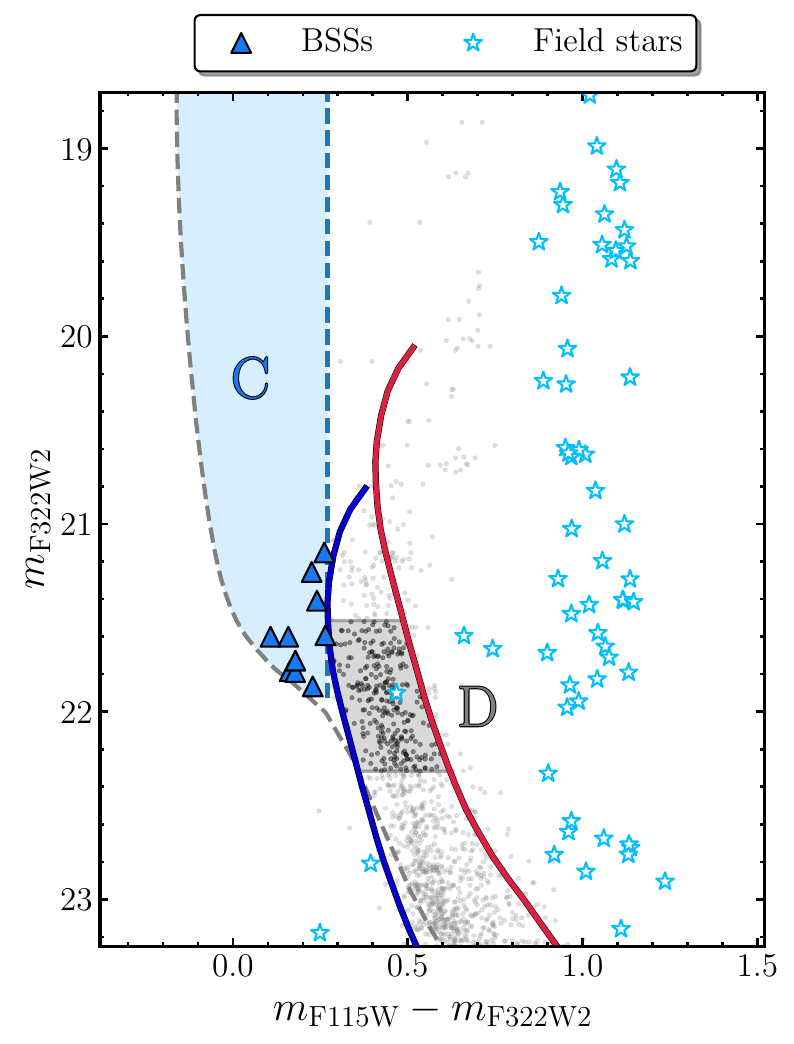}       
    \caption{Identification of candidate BSSs in the field of the SMC. The gray dashed line indicates the zero-age MS isochrone, while the azure dashed vertical line marks the color of the MS turn-off. The blue and red solid line represent the fiducial sequence of MS stars shifted blueward by five times the color error, and the equal-mass binary fiducial, respectively. The azure-shaded area indicates region C, whereas region D is shown in gray. Candidate BSSs are highlighted with blue triangles, and field stars are shown as azure starred symbols. See the text for details.}
    \label{fig:BSSs}
\end{figure}

BSSs are observed as an extension of the MS in a region brighter and bluer than the turn-off in the $m_{\rm F322W2}$ versus $m_{\rm F115W}-m_{\rm F322W2}$ CMD. To identify them, we applied a selection method similar to \citet[]{cordoni2023}. Specifically, we defined two regions in the CMD, hereafter region C and region D, which are shaded in azure and gray in Fig.~\ref{fig:BSSs}, respectively. Region C corresponds to the area bounded by the zero-age MS isochrone (gray dashed line), the MS fiducial shifted blueward by five times the color uncertainty (blue solid line), and a vertical line at the color of the MS turn-off (azure dashed line). All stars located within this region were classified as candidate BSSs, resulting in the identification of 10 BSSs in the SMC field, marked as blue triangles in Fig.~\ref{fig:BSSs}.

For region D, we adapted the definition by \cite{cordoni2023} to the $m_{\rm F322W2}$ versus\,$m_{\rm F115W}-m_{\rm F322W2}$ CMD. Specifically, we identified region D as the interval included between the MS turn-off and one $G_{\rm RP}$ magnitude below the MS turn-off, which, based on the best-fit isochrone, corresponds to 0.8 F322W2 magnitude. Its left boundary is set by the MS fiducial line shifted leftward by five times the color uncertainty, while the right boundary follows the fiducial of equal-mass binaries (red solid line).

The fraction of BSSs in the field of the SMC was estimated using the ratio:
\begin{equation}
    f_{\rm BSS}=\frac{N^C_{\rm obs}-N^C_{\rm field}}{N_{\rm obs}^D-N_{\rm field}^D}
\end{equation}
where $N_{\rm obs}^C$ and $N^D_{\rm obs}$ represent the number of observed stars in regions C and D of the CMD, respectively, while $N^C_{\rm field}$ and $N^D_{\rm field}$ are the corresponding quantities measured for field stars. 

The resulting BSS fraction in the field of the SMC is $f_{\rm BSS}=0.04\pm0.01$, similar to the values measured by \cite{cordoni2023} in Galactic OCs, which range from 0.00 (no BSSs) up to 0.07. Alternatively, adopting the definition of region D from \cite[][see their Sec.\,3]{sollima2008} we measured a lower BSS fraction, $f_{\rm BSS}=0.02\pm0.01$. Despite this difference, both measurements are compatible with those found in other low-density stellar environments, where mass transfer in binary systems is considered the dominant formation channel for BSSs \citep[e.g.,][]{momany2007,cordoni2023}.

\begin{figure*}
    \centering
    \includegraphics[width=.95\textwidth]{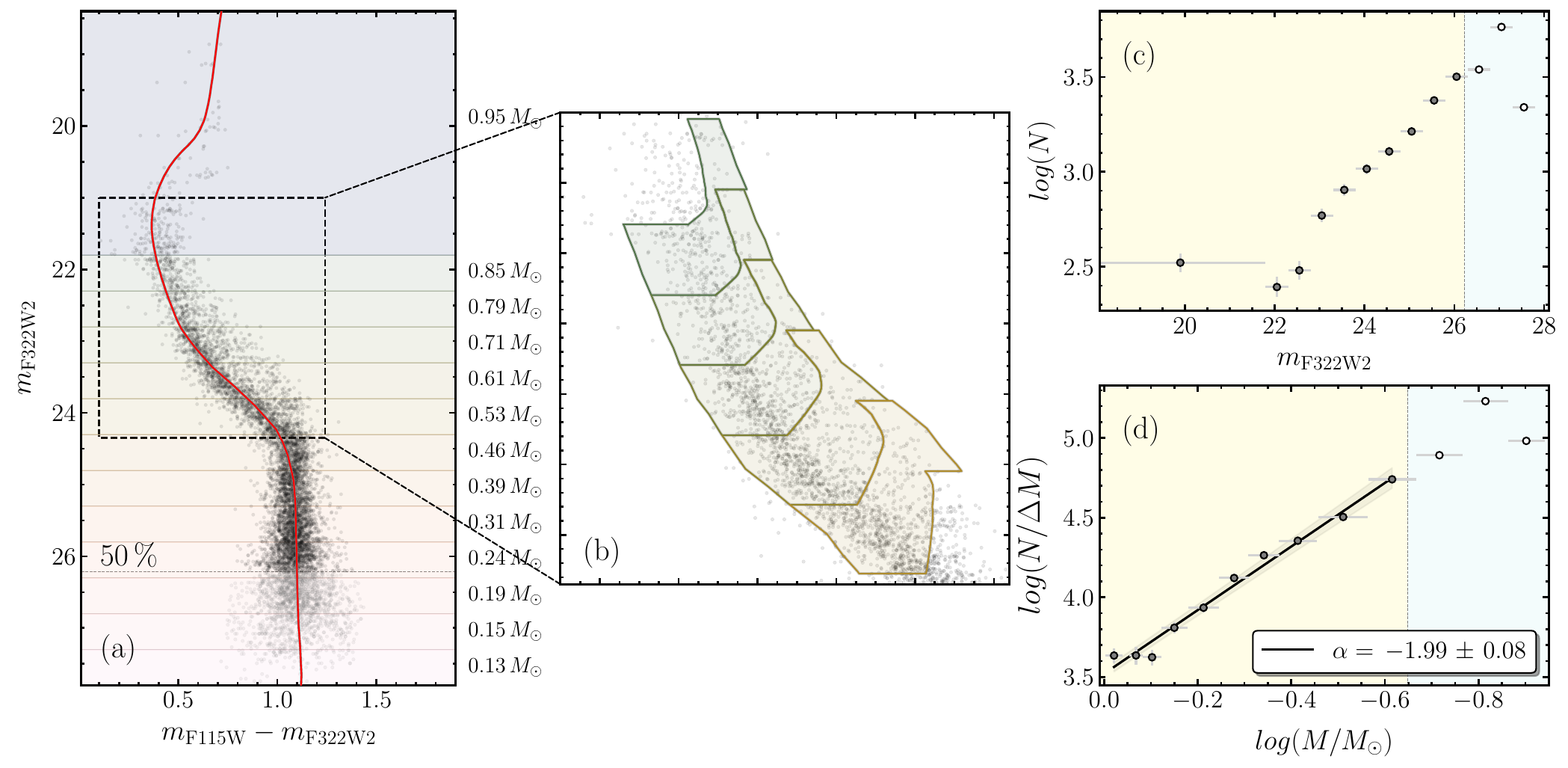}       
    \caption{Determination of the SMC MF. \textit{Panel a.} $m_{\rm F322W2}$ vs.\,$m_{\rm F115W}-m_{\rm F322W2}$ CMD for stars in the SMC. The magnitude intervals are highlighted in different colors, with the corresponding average stellar masses labeled on the right. The red solid line shows the isochrone that best fits the observed stellar population. \textit{Panel b.}  Zoom-in of the CMD focusing on the magnitude range $21.0 < m_{\rm F322W2} < 24.3$. Colored regions indicate the bins used to compute the MF along the upper MS, tailored to include binary stars. \textit{Panels c and d.} The $m_{\rm F322W2}$ luminosity function (c) and corresponding MF (d) for SMC field stars. White points mark bins with photometric completeness below 50$\%$. In panel d, the black solid line represents a linear fit to the observed MF, restricted to regions of the CMD where the photometric completeness exceeds 50$\%$. The derived MF slope is indicated in the bottom-right corner of the figure. In all panels, the gray dashed line indicates the 50$\%$ completeness limit.} 
    \label{fig:SMC_MF_det}
\end{figure*}

\section{The initial mass function of field stars in the Small Magellanic Cloud}
\label{sec:MF}
In the following, we derive the MF of field stars in the SMC. We begin by constructing the observed luminosity function, accounting for the presence of binaries (Sec.~\ref{subsec:SMC_LF}). In Sec.~\ref{subsec:SMC_IMF}, we convert luminosities into stellar masses using the best-fit isochrone to obtain the MF. Finally, we determine the slope of the MF and discuss its implications for the universality of the IMF.

\subsection{The luminosity function of field stars in the Small Magellanic Cloud}
\label{subsec:SMC_LF}
To derive the luminosity function of stars in the SMC field, we adapted the method introduced by \cite{milone2012a} to our dataset. Panels a and b of Fig.~\ref{fig:SMC_MF_det} show the binning scheme used for the analysis. To properly account for binary contamination, we adopted different strategies for the upper and lower MS. Specifically, in the upper MS, we divided the CMD into five regions following a procedure similar to that described in Sec.~\ref{subsec:bin}. The key difference is that, for each of the five magnitude intervals, we only defined region A, namely the area enclosed by the green solid line in Fig.~\ref{fig:bin}, which includes both single MS stars within a fixed 0.5 mag interval and binaries whose primary falls within the same range. Since binary identification in the $m_{\rm F322W2}$ versus\,$m_{\rm F115W} - m_{\rm F322W2}$ CMD is only reliable in the upper MS, we applied a uniform binning scheme with 0.5 mag-wide bins below the MS knee.

In both the upper and lower MS, observational errors cause contamination between adjacent bins, as stars may scatter into neighboring regions. As a result, the total number of stars observed in each region (corrected for completeness), $N_{i}$, can be expressed as:
\begin{equation} \label{eq:LF}
    N_i=\sum_{k}n_kc_{k,i}
\end{equation} 
where $n_k$ is the true number of stars in the $k$-th bin, and $c_{k,i}$ is the fraction of those stars scattered into the $i$-th bin due to observational uncertainties. The contamination matrix $c_{k,i}$ was computed using synthetic CMDs generated with ASs. The true star counts $n_k$ were then obtained by solving the linear system defined in Eq.~\ref{eq:LF}.

To account for the presence of unresolved binaries, we first used the binary fraction with $q<0.6$ measured in Sec.~\ref{subsec:bin} to infer the total binary fraction, $f_{bin}^{\rm TOT}$. Assuming a flat mass-ratio distribution over the range $0<q<1$, we derived $f_{bin}^{\rm TOT}=0.4$. In each magnitude bin, we then randomly assigned binary companions by drawing mass ratios from a uniform distribution between $q=0$ and $q=1$. The corresponding companion masses were converted into F322W2 magnitudes using the mass–luminosity relation from \citet{pietrinferni2021}, and their flux contributions were added to the primary stars in the respective bins.

The resulting luminosity function is shown in panel c of Fig.~\ref{fig:SMC_MF_det}. White points indicate bins with completeness below 50$\%$, and error bars represent Poisson uncertainties.

\subsection{The mass function slope in the field of the Small Magellanic Cloud}
\label{subsec:SMC_IMF}
The red line overlaid on the SMC sequence in panel a of Fig.~\ref{fig:SMC_MF_det} corresponds to the best-fitting isochrone from the BaSTI stellar evolution database \citep{pietrinferni2021}. This fit adopts an age of 7 Gyr, a metallicity of $\rm [Fe/H] = -1.1$, a distance modulus of $(m-M)_{0} = 18.90$, and a reddening of $E(B-V) = 0.05$. Using this isochrone, we converted the observed luminosities into stellar masses, as illustrated in the same panel. The distinct mass intervals used to construct the SMC MF are shown in different colors, with the average mass of stars in each bin labeled accordingly.

The resulting MF is shown in Fig.~\ref{fig:SMC_MF_det}d, where we plot the logarithm of the number of stars per unit mass ($\log(dN/dM)$), normalized by the bin width $\Delta M$, as a function of the logarithm of stellar mass. A previous study by \citet{kalirai2013a}, based on ultra-deep {\it HST} data, probed the MF of the SMC field down to $M=0.37\,M_{\odot}$. In comparison, our {\it JWST} observations extend significantly deeper. According to our mass-luminosity relation, the faintest SMC stars detected in our data, at $m_{\rm F322W2} = 27.9$, correspond to a mass of $M=0.11\,M_{\odot}$. However, the 50$\%$ completeness limit occurs at $m_{\rm F322W2} = 26.22$, which translates to $M=0.22\,M_{\odot}$. This represents a substantial improvement in the ability to study the low-mass end of the SMC MF.

The present-day MF of the SMC field, expected to closely resemble the IMF given the dynamically mixed nature of field stars \citep[][and references therein]{geha2013}, is commonly described by a power-law:

\begin{equation}
\frac{dN}{dM}=k \cdot M^{-\alpha}
\end{equation}

\noindent where $k$ is a normalization constant, and $\alpha$ is the slope. Taking the logarithm of both sides gives the linear form:

\begin{equation}
\log\left(\frac{dN}{dM}\right) = \log(k) - \alpha \cdot \log(M)
\end{equation}
We performed a least-squares linear fit to the observed $\log(dN/dM)$ versus $\log(M)$ distribution, obtaining a best-fitting slope of $\alpha = -1.99 \pm 0.08$ (black solid line).  To validate this result, we also estimated the MF slope using the Maximum Likelihood Estimation (MLE) approach, which assumes a power-law distribution with a lower mass limit $M_{\rm min}$ and an infinite upper bound. Applying this method to the observed mass distribution yields $\alpha_{\rm MLE} = -1.85 \pm 0.01$, a value consistent with our least-squares estimate within $2\sigma$.

Several studies have suggested that the stellar MF is better described by a broken power law. In particular, \citet{kroupa2001} reported a change in slope from $\alpha = -2.3$ to $\alpha = -1.3$ at $M = 0.5\,M_{\odot}$. When we imposed a break at $M = 0.5\,M_{\odot}$, we obtained $\alpha = -1.69 \pm 0.15$ for $M < 0.5\,M_{\odot}$ and $\alpha = -2.02 \pm 0.29$ for $M > 0.5\,M_{\odot}$. The high-mass slope agrees with the Kroupa value of $\alpha = -2.3$, while the low-mass slope is slightly steeper than the $\alpha = -1.3$ measured in the Galactic field, suggesting a relative deficit of low-mass stars in the SMC. Consistent with the findings of \cite{kalirai2013a}, however, we found that a single power law reproduces the SMC field MF well and no evidence for a broken MF from our data only. Moreover, the broken power-law fit is statistically consistent with a single power law of slope $\alpha =-1.99$ at 1$\sigma$ level.  Hence, in the following, we will consider the value obtained from the entire mass range.

\subsection{Comparison with previous studies}
\label{sub:MFcomp}
\begin{figure}
    \centering
    \includegraphics[width=.45\textwidth]{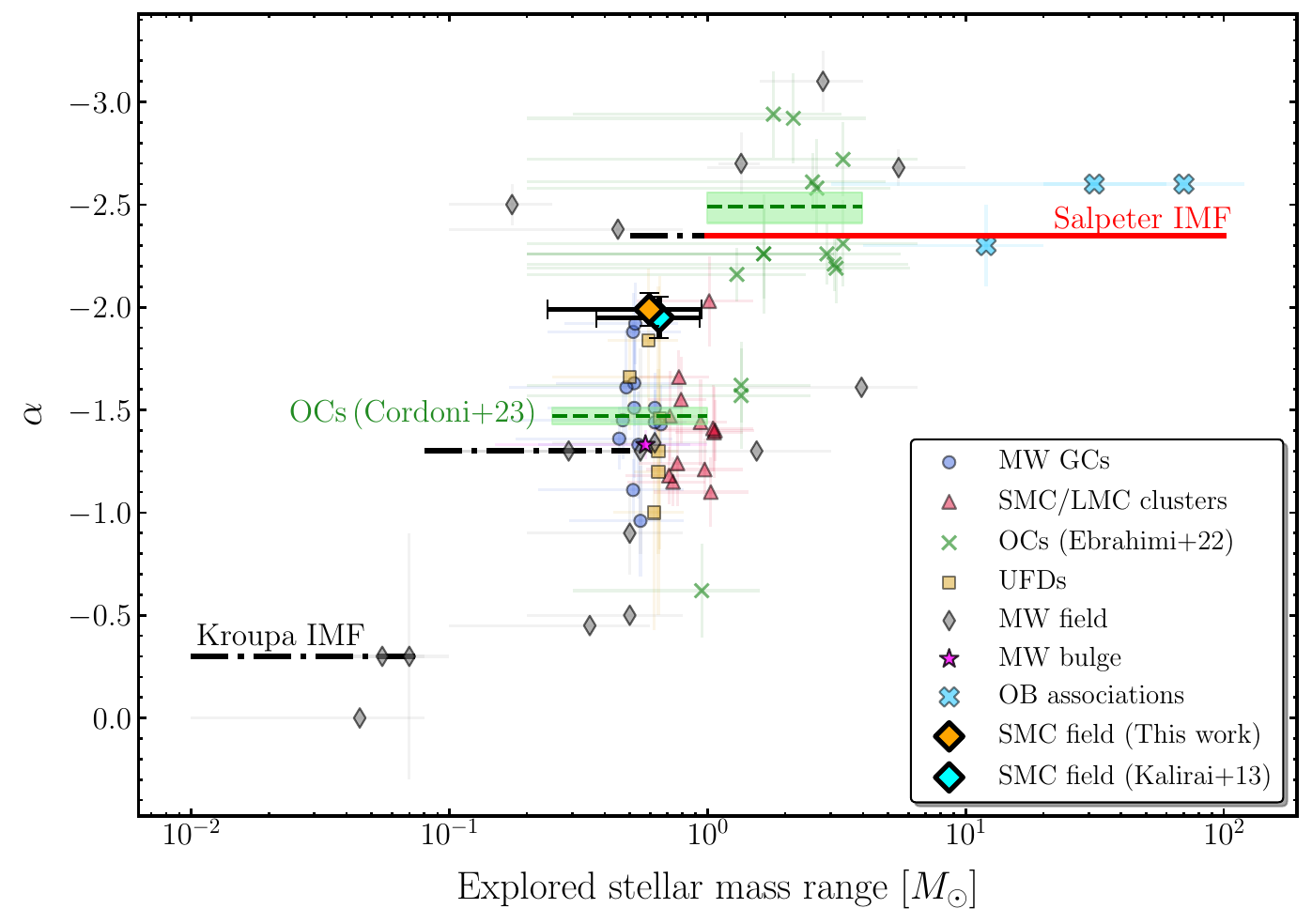}       
    \caption{MF slope, $\alpha$, as a function of the explored stellar mass range, compiled from a variety of Galactic and extragalactic environments. Each point represents the best-fit power-law slope over a specific mass range, with horizontal error bars showing the width of the corresponding mass interval. Blue circles denote Milky Way GCs, crimson triangles correspond to SMC/LMC clusters, green thin crosses to OCs, gold squares to UFDs, gray diamonds to Milky Way field stars, the magenta star to the Galactic bulge, and cyan thick crosses to OB associations. The orange diamond shows the measurement for the SMC field from this work, while the nearby cyan diamond corresponds to the result by \citet{kalirai2013a}. The solid red line marks the canonical \citet{salpeter1995} IMF slope, while the dashed black lines represent the segmented IMF slopes from \citet{kroupa2001}. The green dashed line corresponds to the slopes measured by \cite{cordoni2023} for Galactic OCs.}
    \label{fig:alpha_plot}
\end{figure}

The MF slope, $\alpha$, measured for the SMC field  is shown in Fig.~\ref{fig:alpha_plot} alongside a broad compilation of MF slopes derived from various Galactic and extragalactic stellar populations. These include Galactic GCs \citep[blue circles;][]{baumgardt2023}, SMC/LMC clusters \citep[crimson triangles;][]{baumgardt2023}, OCs \citep[green thin crosses and green dashed lines;][]{ebrahimi2022,cordoni2023}\footnote{As in \cite{baumgardt2023}, we only show clusters with lifetimes and relaxation times larger than their ages. For these objects, the present-day MF should still reflect the initial one.}, UFDs \citep[gold squares;][]{geha2013,gennaro2018a}, the Milky Way field \citep[gray diamonds;][]{reid1999,reid2002,schroder2003,kroupa2002,allen2005,metchev2008,pinfield2008,bochanski2010,sollima2019}, the Galactic bulge \citep[magenta starred symbol;][]{zoccali2000}, and OB associations \citep[cyan thick crosses;][]{reyle2001,schultheis2006,vallenari2006}.

Our SMC field measurement (orange diamond) extends down to $0.22\,M_{\odot}$ at the 50$\%$ completeness limit, probing significantly lower stellar masses than most prior extragalactic studies. This enables a robust characterization of the MF in a low-metallicity, dynamically mixed stellar population. The resulting slope, $\alpha = -1.99 \pm 0.08$, is in excellent agreement with the value reported by \citet[][cyan diamond]{kalirai2013a}, $\alpha = -1.90^{+0.15}_{-0.10}$, based on ultra-deep {\it HST} data over the mass range $0.37$–$0.93\,M_{\odot}$.

In the context of broader stellar populations, the SMC slope is shallower than the canonical Salpeter value (red solid line). Additionally, it is consistent with the MF slopes observed in several Milky Way populations, including some OCs and GCs. Similarly, the SMC slope agrees with the values measured within SMC/LMC clusters suggesting no strong variation in the MF between field and cluster environments at comparable metallicities.

\section{Summary and Discussions}
\label{sec:concl}
In this work, we used deep {\it JWST} NIRCam photometry in the F115W and F322W2 bands to investigate stellar populations in the field of the SMC. Our study leveraged the unprecedented spatial resolution and depth of {\it JWST} observations to explore key aspects of stellar populations in a low-metallicity and low-density environment. Specifically, we primarily focused on measuring the fraction of binary systems and deriving the MF. These results provide valuable insights into the formation and evolution of stars in dwarf galaxies and contribute to ongoing discussions about the universality of the IMF. To properly account for binary contamination in the determination of the MF, we first concentrated on binaries:
\begin{itemize}
    \item We determined for the first time the fraction of unresolved binaries with mass ratios $q>0.6$ finding $f_{\rm bin}^{q>0.6}=0.14\pm0.01$. This value is independently confirmed through comparisons with synthetic CMDs accounting for the SMC’s metallicity distribution, yielding a consistent value of $f_{\rm bin}^{q>0.6}=0.14\pm0.03$. 

    \item We investigated the presence of BSSs in the CMD. In particular, we measured a BSS fraction of $f_{\rm BSS}=0.04\pm0.01$. This value is consistent with previous studies of BSSs in low-density stellar environments, such as dwarf spheroidal galaxies and GC outskirts, where dynamical interactions are rare and mass transfer in binary systems is believed to be the dominant formation channel \citep[e.g.,][]{momany2007,ferraro2012}. Notably, \cite{cordoni2023} showed that OCs also exhibit significant BSS populations with similar structural dependencies, reinforcing the idea that mass exchange in binaries is a universal mechanism shaping BSS formation, from sparse stellar systems like OCs to entire dwarf galaxies like the SMC.

\end{itemize}

The binary fraction measured for the SMC field in this work is comparable to that observed in Galactic OCs of similar ages. Furthermore, our result for the SMC field is consistent with the total binary fraction measured for Milky Way field stars of similar primary mass. These findings align with the general trend that binary fractions tend to be higher in lower-density environments, where dynamical interaction rates are relatively low. In contrast, Galactic GCs, characterized by much higher stellar densities and interaction rates, experience more frequent binary disruptions, resulting in systematically lower binary fractions \citep[e.g.,][]{sollima2007, milone2012b, milone2016a}. This comparison suggests that the binary star formation and evolutionary processes in the low-density environment of the SMC resemble those in Galactic OCs and field populations, supporting the notion that environmental density and dynamical interactions play a key role in shaping binary fractions across different stellar systems.

Our main results on the MF of the SMC field can be summarized as follows:

\begin{itemize}
    \item We constructed the luminosity function of SMC field stars and modeled it using synthetic populations that account for both completeness effects and unresolved binaries. Using the best-fit isochrone, we converted magnitudes to stellar masses, allowing us to derive the MF down to $M = 0.22\,M_\odot$, significantly extending previous limits \citep{kalirai2013a}.
    \item We measured the present-day MF of field stars in the SMC down to 0.22 $M_\odot$, finding a power-law slope of $\alpha = -1.99 \pm 0.08$. This result provides an independent confirmation of the findings by \citet{kalirai2013a}, who reported a slope of $\alpha = -1.90$ over a narrower mass range (0.37–0.93 $M_\odot$) in a different SMC field. Our measured slope is significantly shallower than the canonical Salpeter value ($\alpha = -2.35$), challenging the notion of a universal IMF.
\end{itemize}

The shallower MF slope observed in the SMC implies that star formation in low-metallicity, low-density environments favors the formation of relatively more intermediate- to high-mass stars compared to low-mass stars. This suggests that environmental conditions such as metallicity, temperature, and gas density influence the fragmentation of molecular clouds and the resulting stellar mass distribution. Such variations challenge the universality of the IMF and have important consequences for the chemical evolution, feedback, and dynamical history of dwarf galaxies like the SMC.

\begin{acknowledgements}
We thank the anonymous referee for various suggestions that improved the quality of the manuscript. This work has been funded by the European Union – NextGenerationEU RRF M4C2 1.1 (PRIN 2022 2022MMEB9W: “Understanding the formation of globular clusters with their multiple stellar generations”, CUP C53D23001200006) and from INAF Research GTO-Grant
Normal RSN2-1.05.12.05.10 (PI A. F. Marino) of the “Bando INAF per il Finanziamento della Ricerca Fondamentale 2022”. T. Ziliotto acknowledges funding from the European Union’s Horizon 2020 research and innovation program under the Marie Skłodowska-Curie Grant Agreement No. 101034319 and from the “European Union – NextGenerationEU”.
\end{acknowledgements}

\bibliographystyle{aa}
\bibliography{aa56239-25}
\end{document}